\documentclass[pre,preprint,aps,]{revtex4}
\usepackage{graphicx}
\usepackage{comment}
\usepackage{amsmath,amssymb,amsthm} 
\usepackage{bm}
\usepackage{verbatim}
\usepackage{float}

\begin{document}

		\title{ Dynamics of homogeneous nucleation}

\author{S\o ren Toxvaerd$^{a)}$ }
\affiliation{DNRF centre  ``Glass and Time,'' IMFUFA, Department
 of Sciences, Roskilde University, Postbox 260, DK-4000 Roskilde, Denmark}
$^{a)} st@ruc.dk$
\date{\today}

\vspace*{0.7cm}

\begin{abstract}
The classical nucleation theory  for homogeneous nucleation is  formulated as a theory
 for a density fluctuation in a supersaturated gas at a given temperature. But Molecular Dynamics  simulations
 reveal that it is  small cold clusters which initiates the nucleation. The temperature
in the nucleating clusters fluctuate, but the mean
temperature  remains below the temperature in the supersaturated gas until they reach 
the critical nucleation size.
 The critical nuclei have, however,  a temperature equal to the supersaturated gas.
The kinetics of homogeneous nucleation is not only caused
by a grow or shrink by accretion or evaporation of
 monomers only, but by an exponentially declining
change in cluster size per time step equal to 
 the cluster distribution in the supersaturated gas.

\end{abstract}
\maketitle

\section{Introduction}
 The classical nucleation theory (CNT) \cite{Volmer,Becker} was formulated for 
almost ninety years ago. The theory describes the nucleation of a liquid droplet 
in a supersaturated  gas as the creation of a critical nucleus,  where the gain in free energy of the liquid in the nucleus by
increasing the number of particles  equals the cost of an increased  surface free energy. Since then there has been
  continuous refinements  of the
basic ideas of CNT \cite{Lothe,Reiss,Kalikmanov}, and formulations of semiphenomenological models \cite{Dillmann,Viisanen,Laaksonen,Reiss1}.

 Parallel with the development in the theories and the experimental investigations of nucleation, the development in
computer capability has make it possible to determine homogeneous nucleation 
 by various simulation methods \cite{Daan,Matsumoto,Tox1,Diemand}. The
 Molecular Dynamics (MD)  simulations make it possible directly
 to obtain detailed information,  not only about  the  thermodynamics, but also the kinetics
at homogeneous nucleation. 
  The  present simulations
reveal that the creation of a critical nucleus  always is initiated by a small $\textit{cold}$ cluster,
 much smaller than the  the critical nucleus, and the cluster remains colder than
 the supersaturated gas until the critical nucleus is created. The  dynamics and the properties of
the nucleating cluster
 deviate at several points from the assumptions in the traditional CNT and its refinements.

The article is organized as follows. Section II describes  the MD   simulation of homogeneous nucleation,
and the data collection of the properties of the nucleating clusters are given in  subsections (IIa, IIb). The properties of the
nucleating clusters are presented in Section III, which first describes the supersaturated state in subsection IIIa.
A short summary   is given in Section IV.  

\section{ Molecular Dynamics  simulation of homogeneous nucleation }

The system consists of $N=40000$ Lennard Jones (LJ)
particles in a cubic box with periodical boundaries \cite{ToxMD}. The MD simulations
are performed with the central difference algorithm in the leap-frog
 version, and the forces for pair interactions  greater than $r_{cut}$ are ignored. There are different ways
to take the non-analyticity of the potential at $r_{cut}$  into account. The most stable and energy conserving way
is to cut and shift the forces (SF) \cite{Tox2}, but most simulations are
 for cutted and shifted potentials (SP). The extensive MD simulations
of homogeneous nucleations \cite{Diemand,Angelil} were performed for a SP cut with $r_{cut}= 5 \sigma$,
 and with the conclusion, that even the very weak attractions behind this long range interaction increases the
nucleation rate a little. The  present simulations are for SF with  $r_{cut}= 5 \sigma$. We have
investigated   ensembles of 25 simulations with SP and SF with $r_{cut}= 5 \sigma$ and found no significant differences
in the obtained data,   inclusive the temperatures of the nucleating cluster,  within the statistical uncertainties.

The discrete MD is exact in the sense that there exist a "shadow Hamiltonian" for sufficient small time steps for which
the discrete MD positions lie on the  trajectories for the Newtonian dynamics with the shadow Hamiltonian \cite{Tox3}. The proof is
based on an analytic expansion, but it is not possible to determine
 the range of convergence of the expansion. Numerical investigations indicate, however, that the energy is conserved
 for time steps,  $\delta t$,  significant bigger than traditional used in MD \cite{Tox4}. The  simulations
in  \cite{Diemand,Angelil}  were performed for a time increment $\delta t=0.01$, well inside
the estimated convergence radius. We have performed ensemble simulations with $\delta t=0.01$ and
$\delta t=0.0025$ and found no significant differences in the data ( $ T_{cl}, n_{coor.no.},u_{pot}$)
 for the  nucleations within the statistical uncertainties.

 The time averages of the MD simulations correspond to the microcanonical averages with
a constant energy (NVE). The dynamics can, however, be constrained to a given mean
 temperature (NVT), e.g by a Nos\'{e}-Hover constraint \cite{Tox5}.
 As concluded in \cite{Wedekind} and \cite{Diemand}, a temperature constraint
does not affect the homogeneous nucleation in the very big particle systems. We have performed
 NVE and NVT ensemble simulations and have reached to the same conclusion. The
 nucleation data in Table 1 and Table 2 in  Section III are for nucleations  at
the temperature  $T=0.80$ and for five  different  densities, and for ensemble simulations of 25 independent
NVT simulations  at each density with a time increment  $\delta t= 0.01$ and with the  Nos\'{e}-Hoover thermostat with a friction
$\eta=0.05$. 

\subsection{MD simulation of homogeneous nucleation}

    A  supercooled gas can remain in the quasiequilibrium state (QES) for some times, but  sooner or later  it will
 form  some critical nuclei  and separate into  a gas with liquid droplets.
Homogeneous nucleation with MD is a stochastic and chaotic process.
 It is not possible to predict, when a nucleation takes place,
and the round-off errors of the floating point data for the chaotic MD dynamics make it
difficult to reproduce the dynamics  of the nucleation in  detail.
 For these reasons we have (for the first time)  determined the cluster distribution and their properties
 directly in the system $\textit{at each time step  during the simulation}$
  and thereby collect the necessary
information about the kinetics  and the temperature of the  nucleating cluster directly during the nucleation.

 The dynamics and the
  energies, temperatures and coordination number of the nucleating clusters are determined for
 the QES densities $\rho=$0.027, 0.0275, 0.028, 0.0285 and 0.029 along the isotherm $T=0.80$.
 But although  the nucleating cluster
exhibits some rather well defined behaviour, it varies, however, for the different nucleating clusters.
For this reason  ensembles of independent nucleating clusters are performed, and the results in Table 1 and Table 2
 are the means of 25
 independent nucleations, and with the uncertainties obtained from the standard deviations of the 25 independent nucleations at
each density.

It is necessary to determine the distribution of  clusters   every
time step in order to obtain  the dynamics of nucleation.
 This can be done without a significant increase in computer time.
Normally a MD program for a  system of many thousand  particles  is optimized by  sorting the particles in  pairs of particles,
which interact or might interact within a short time interval. The present MD is performed
by a double sorting, first by sorting the particles into
subboxes of side length $r_{cut}+\delta$, following by a neighbour linked-list sorting
of pairs of interacting particles within subboxes and
their neighbour subboxes. The list is then updated when one of the $N$ particles have moved more
than a distance $\delta/2$. The computer time increases almost linearly with
the number $N$ of particles in the system by this double sorting of nearest neighbours whereby one can simulate big systems.

The cluster distribution can be obtained from the list of pairs of neighbour particles 
 at the same time as the forces between the particles is determined.
The force between pairs of particles, no. $i$ and $j$ is calculated for
  $r_{ij} \leq 5 $. At this point in the program one also notices if the distance
is smaller than the binding distance  $r_{ij} \leq  r_c$.
The total distribution of clusters are obtained in the following way:
At a given time, and before the forces are calculated, all the particles are given  a cluster number  equal to zero
(in an array  $clusterno(N,0) $).
They are not yet associated to any cluster.
The MD program starts the
force calculation by calculating the distance for the first pair of neighbours, and  if  $r_{ij} \leq  r_c$
  these two particles  are associated to cluster number 1 (i.e.  $clusterno(i)=1,$  $  clusterno(j)=1  $), and
  the cluster number 1  contains (in a corresponding array, $ cluster(1,1)=i$ and $cluster(1,2)=j$)
  the information that  particle number $i$ and $j$ is associated to this cluster.
 A new pair of neighbors is investigated and
  it happens that $i$ or $j$ already are associated to a  cluster ( $clusterno(i) \neq 0$ or $clusterno(j)\neq  0 $), then $j$ is included
in $i$'s cluster or $\it{vice} $ $\it{verse}$, or it happens that both particles are within two
different clusters, which then are merged together.   The full
determination of all the clusters and  with particle numbers is then completed at the end of the force calculations. The computer time
 increases linearly with $N$, and it is possible to obtain the cluster distribution every time steps for very big MD systems
without a significant increase in computer time
 \cite{Laradji}.

\subsection{Determination of cluster properties}
The particles in a cluster are within a "binding" distance,  $r_c$,  to some other particles in the cluster.
Traditionally one chooses
this distance to be significant smaller than the range of the first coordination shell \cite{Tox6} in
order to ensure that  only particles which  are energetically  tied together
are included in the cluster. In the present investigation we have used  $r_c=1.4$.
The precise definition of how many particles a given particle is close to- and how close to it is, is not
 important for the determination of a cluster and critical nucleus .  For details concerning
the definition of a cluster  see e.g. \cite{Daan,Tox1}. The  size of the  critical nucleus
depends, however, on the choice of the coordination number.

The particles within a cluster are here characterized by that  they have
  coordination numbers, $n_{co.no.}$, greater than one, i.e a particle is at least near two other particles.
By this definition particles within the "corona" of the cluster are included whereas a particle
which just collide with a particle at the "surface" of the cluster is not.
 An example of the time evolution of the number of
particles, $n_{cr}(t)$, in  a nucleating  cluster is shown in Figure 1. A small cluster of $\approx 15$
particles grow to a critical nucleation size at time $t \approx 6050$ (inset). The time evolution of 
  $n_{cl}(t)$ is obtained  
for three different  definitions. The red curve is $n_{cl}(t)$ for a nucleating cluster where all particles
have at least one neighbour particle in the cluster.  The green curve is $n_{cl}(t)$ with the definition used here, where
a particle has at least two other neighbours. In \cite{Daan} it is demonstrated that all LJ particles in a  liquid  
have at least five nearest neighbours, and the blue curve shows $n_{cl}(t)$ with this definition.
The time evolutions of  $n_{cl}(t)$  are, however synchronous for the three criteria, and the obtained data for the homogeneous
nucleation, except the size of the  critical nucleus, depend not on the precise  definition of a cluster.

\begin{figure}[!]

\includegraphics[width=7.5cm,angle=0]{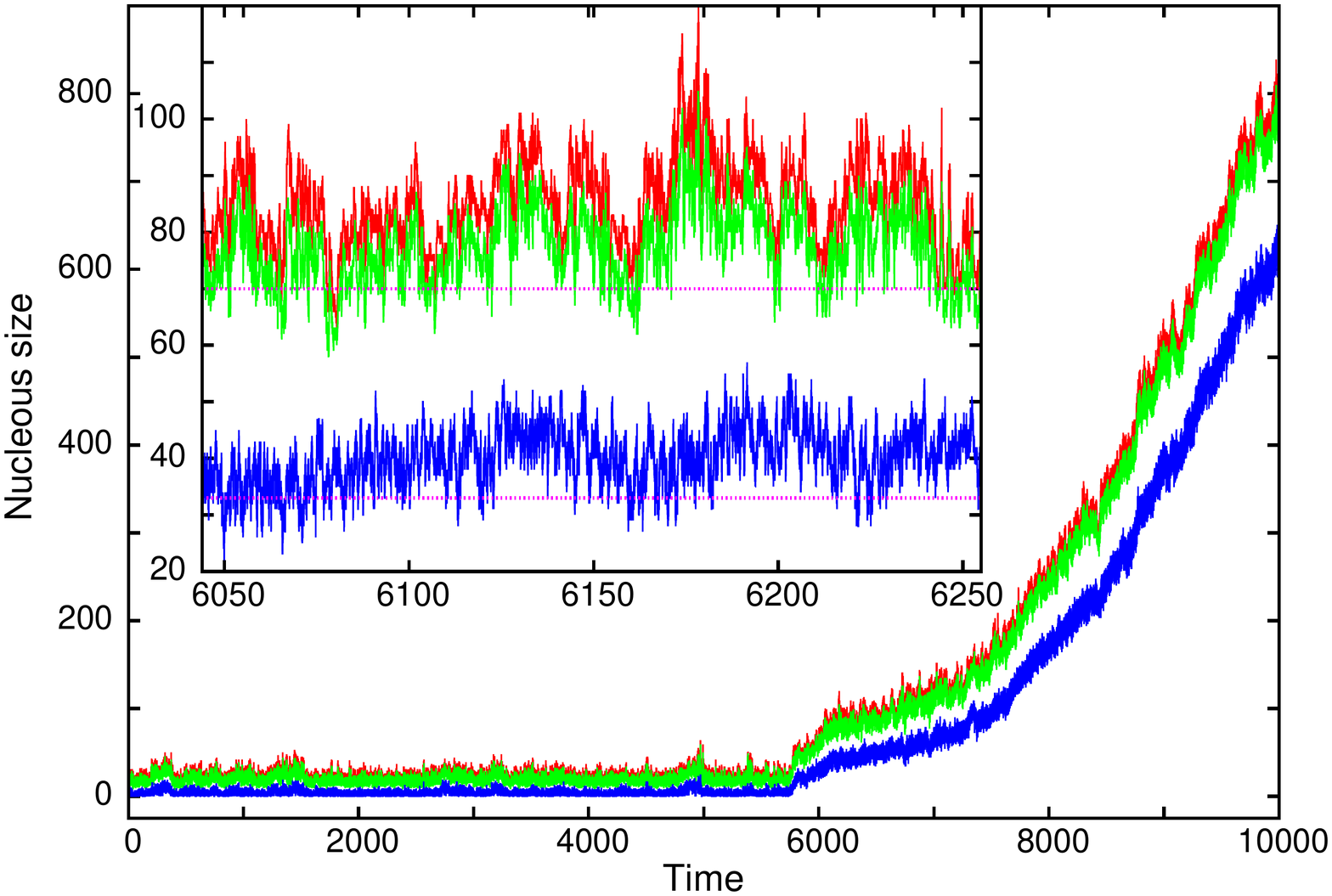}
\caption{ Time  evolution of the number, $n_{cl}(t)$, of particles
 in the nucleating cluster at the super cooled
state $T=0.80$, $\rho=0.0275$.
The evolution are for three different cluster criteria: The red curve is the number of  particle with at least
one neighbour particle; green is the number of particles with at
least two neighbours, and blue is the number of
particles with at least five  neighbour particles. The inset shows the evolution in the transition zone at the critical
nucleation barrier,
and the straight (magenta) dashed lines mark the corresponding critical nuclei sizes.}
\end{figure}

The temperatures, densities, energies, etc. are obtained from the positions and velocities of the particles in the first
 nucleating
cluster.
But there is a practical problem. The nucleation is a stochastic and chaotic process, and it is not possible to determine in advance
which of the small clusters  in the supersaturated gas which  begin to nucleate.
 In practice it requires a very complicated and extensive record of clusters,
if one, in this way shall  identify the  nucleating clusters from the birth to when they reach
 the size of the  critical nucleus. We have overcome this problem   by only recording
the properties of the $\it{biggest}$ $\it{ cluster}$ each time step. The biggest cluster is the first to 
overcome the nucleating energy barrier. But although the nucleation is a  rare event this can, however, lead to errors
 if the recorded start of the growth of the biggest cluster
is for another cluster than a  cluster which shortly after   overhauls and nucleates.
 This event, which happens in a few cases is, however, easily
 identified from a jump in the recorded position of the center of mass of the biggest cluster during the nucleation, which indicates that
the record of size and data is shifted to another cluster. The
simulation is then rejected, and a new independent nucleation is performed.
By this procedure we obtain  the complete  data  for the first nucleating cluster in the system.

The instant temperature, $T(t)$, of the nucleating cluster is determined
in the usual way from the kinetic energy, $E_{kin}(t)$, of the particles
in the cluster at time $t$ as $T(t)=2/3E_{kin}(t)$. (For a determination and discussion of the temperature in small clusters and
 in a non-equilibrium system
see \cite{Wedekind}, \cite{Hoover}.) 
 The density at the center of a cluster is determined indirectly from the coordination number
of the particle closest to the centre of mass of the
cluster.  In the LJ system with spherical symmetrical potentials the mean coordination number of particles
within a given distance gives a precise measure of the
mean density  in the uniform fluid. The question is: when is the density
of the corresponding liquid established at the center of the nucleating cluster? The present simulations are performed
for  densities along the isotherm $T=0.80$, where the corresponding
liquid density is $\rho_l=0.7367$ \cite{Watanabe} and where bulk liquid particles with a sphere with $r_c=1.4$
 have a mean coordination  $n_{coor. no}(l)=  9.02$ and a mean potential energy $u=-5.116$. The data for the corresponding gas
with density  $\rho_g=0.0174$ are $n_{coor. no}(g)= 0.33$ and $u=-0.186$.

The connection between the  coordination number  $n_{coor. no}$  and the density  at
the center of the cluster  is established by determining the coordination number
 for particles in an uniform fluid  for different densities.  There is, however a problem since
a fluid with a density in between the corresponding densities $\rho_g$ and  $\rho_l$ will phase separate.
This problem is circumvented
 by performing  MD where only the forces from particles within the first coordination shell are
included in the dynamics \cite{Tox6}. A system with only these  forces has a radial distribution similar to the LJ system,
and the exclusion of the long attractions in the dynamics ensures that the uniform distribution is maintained without
phase separation \cite{Tox7}.
 The mean coordination numbers for  a particle in  these uniform fluids at $T=0.80$ and with the mean
densities $\rho$= 0.70, 0.65, 0.60 and 0.5 are $n_{coor. no}$=8.40 for $\rho$= 0.70;  $n_{coor. no}$=7.66 for $\rho$= 0.65;
 $n_{coor. no}$=6.94 for $\rho$= 0.60 and  $n_{coor. no}$=5.53 for $\rho$= 0.50. These data are used to determine the
density at the center of the nucleating clusters by interpolation.

\section{ Dynamics of homogeneous nucleation }

\subsection{The quasi-equilibrium state (QES)}
A QES state with supersaturations $S$ is caraterized by that the pressure $ P(\textrm{QES})$ is higher than the pressure $P_l=P_g$
in the corresponding equillibrium system of liquid and gas.
The five supersaturated states correspond to moderate supersaturations $S$ ($ S \equiv P(\textrm{QES})/P_l \approx \rho(\textrm{QES})/\rho_g \approx
1.6$).
 The supercooled  QES are characterized by that they only contain clusters with
cluster sizes significant smaller than the size of the critical nucleus.
The clusters are identified at each time step and the biggest cluster is investigated in details. 
 The biggest cluster contains in mean $n_{cl}  \approx 10-20$ particles for  the density states at the isotherm
$T=0.8$,  whereas the critical cluster size  $n_{cl}$ is $  \approx 70$  (Figure 1). 
The biggest cluster in the QES occasionally reach a size of  $n_{cl}  \approx 70$
within a fluctuation, but the nucleation appears only for a cluster which exceeds  $n_{cl} = 70$.

The  mean distribution of clusters in the QES state  is shown in Figure 2 (blue line) together with
the change in cluster size, $\delta n_{cl}(t)$,  per time step of the biggest cluster in the QES (red line) and
at nucleation (green filled circles).
The distributions in the figure show that the kinetics of homogeneous nucleation, given by  $\delta n_{cl}(t)$, is not only caused
by a grow or shrink by accretion or evaporation of
 monomers only, as in general assumed in classical nucleation theories, but by an exponentially declining
change in cluster size per time step. Secondly we notice that the kinetics  $\delta n_{cl}(t)$ in the QES state
as well as at nucleation, scales well with the
distribution of clusters in the supersaturated gas, so the change in cluster size per time step of the biggest cluster
at nucleation is proportional
with the cluster distribution in the QES, as expected for the kinetics  in an equilibrium state. 

\begin{figure}[!]
\begin{center}
\includegraphics[width=7.5cm,angle=0]{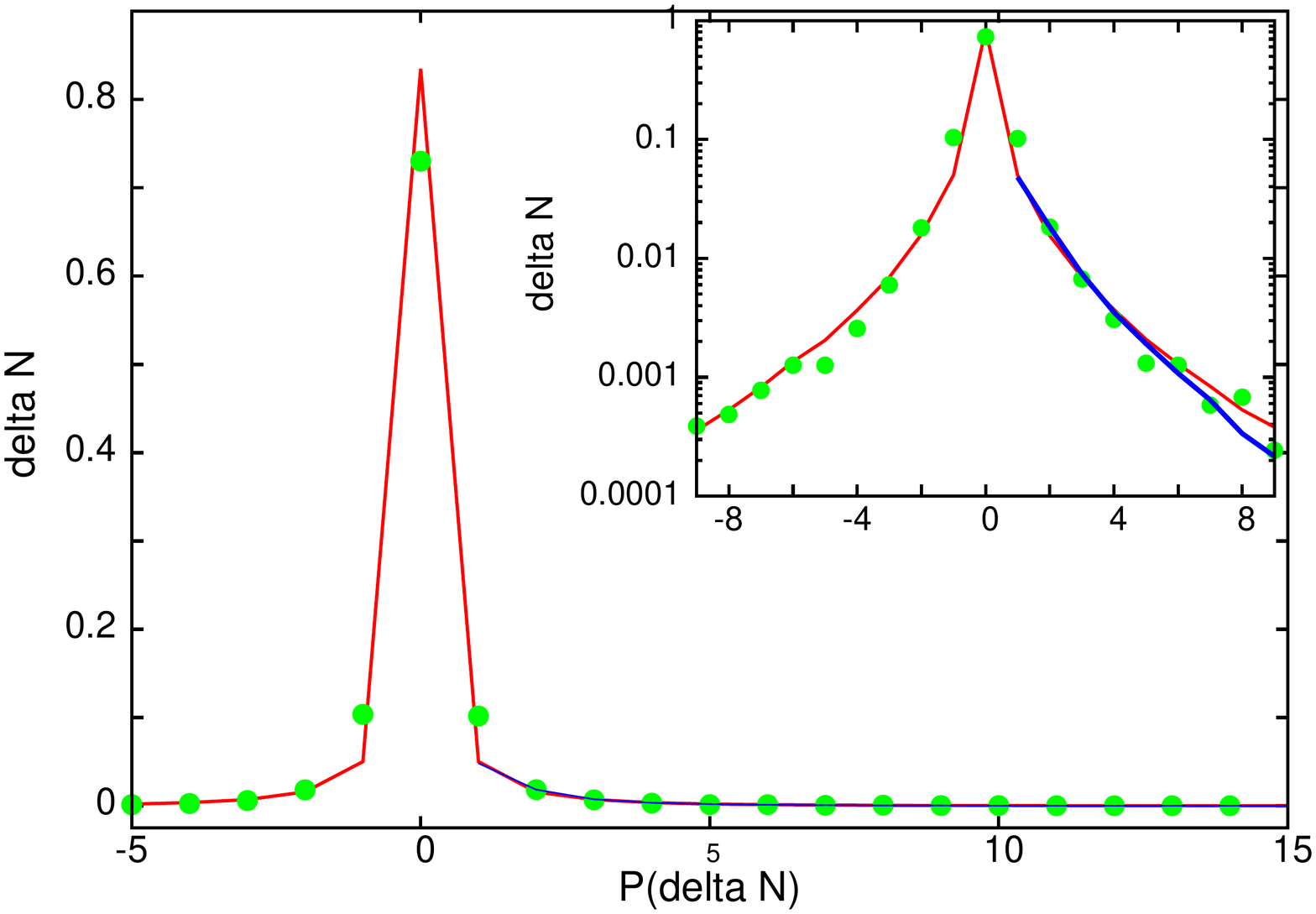}
\caption{ Mean  fraction $P(\delta n_{cl}$) of change in the size  $\delta n_{cl}$ of the
biggest cluster at  $T=0.80$ and $\rho=0.0275$,
 with the fraction  $log P(\delta n_{cl}$) in the
inset. Red line is for the change in the QES state and the green filled circles are for a 
nucleating cluster. The blue  curve is the (scaled) distribution of cluster sizes in the quasi-equilibrium state.
}
\end{center}
\end{figure}

The coordination number,  $n_{co.no.}=5.0_{\pm 2}$,  of the particle nearest the center of mass of the cluster, which
 corresponds to a density $\rho \approx 0.46$,  reveals that the local sphere at the center
is not very liquid-like. (The densities of the corresponding liquid and gas at $T=0.80$
are $\rho_l=$0.7367 and  $\rho_g=$0.0174, respectively \cite{Watanabe}, and with a mean coordination number, $n_{co.no.}(l)$=9.02 in the liquid
 and  a potential energy per particle, $u(l)= -5.116$. The central particle in biggest clusters in the QES has a mean potential energy
 $u= -2.53_{\pm 9}$.) 

\begin{figure}[!]
\begin{center}
\includegraphics[width=5.5cm,angle=-90.0]{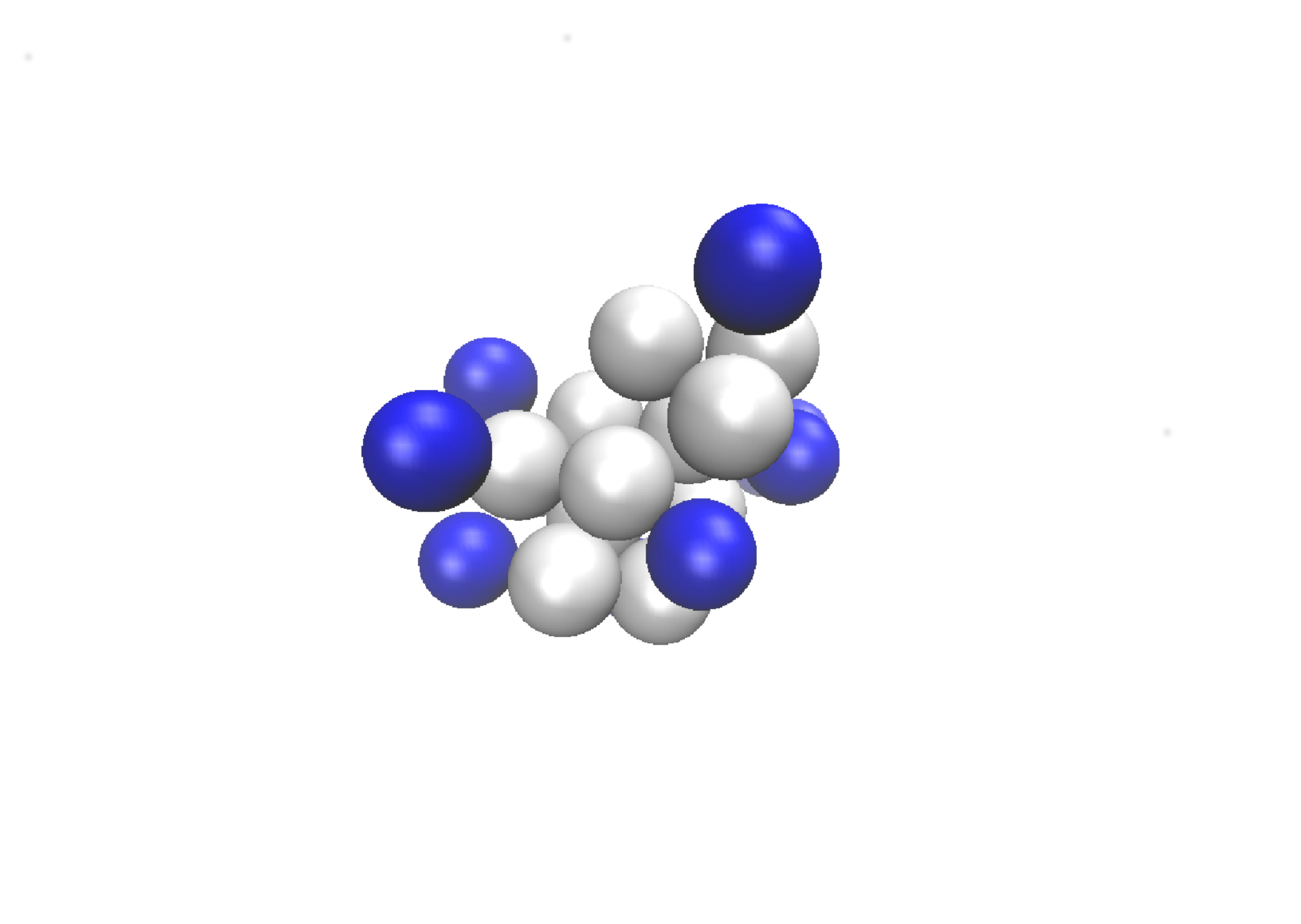}
\end{center}
\caption{ A cold  cluster consisting of 24 particles at the onset of nucleation. The instant temperature is $T_{cl}=0.77$.
 The grey particles are the 14 "liquid like" particles with at least
five nearest neighbours, and  the 10 blue  particles are the weaker attached particles with at least two neighbours. This cluster is shown
in a succeeding figure at nucleation, and later  as  a droplet.
}
\end{figure}

The temperature of an ensemble $n$ of particles at equilibrium fluctuates proportional to $\sqrt{n}$ and 
the temperature in the small clusters in the QES state fluctuates very much. But
 the  mean temperature of the biggest non-nucleating cluster in the QES state
is determined to be $T_{cl}=0.799_{\pm 4}$ for $\rho=0.0275$) and  equal to
 the mean temperature of the system, as must be the case for an equilibrium state.
 This result is, however, rather misleading since it is always $\textit{  a
 cold cluster}$ which initiates the nucleation and nucleate.

\subsection{The nucleation}

 The temperature of the nucleating clusters varies during the nucleating, as also observed by \cite{Wedekind,Angelil}, but
this observation overshadows the fact that the nucleating clusters are cold at the onset of nucleation and remain  cold 
   during the  nucleation.

The result of the ensemble simulations of nucleation for the different densities $\rho$ are collected in Table 1 and Table 2.
Table 1 contains the averages over the time interval $\Delta t_{nucl}$  for the twenty five successful clusters 
with $n_{cl} \approx 10-20$ from  they begin the growth at $t_{nucl. start}$, and  reach the barrier $n_{cl} =70$.
This nucleation barrier is not exact, but we have noticed that a cluster when it reaches
$n_{cl} =70$ either continues the growth,  (Figure 1),
 or in a few cases
shrinks and disappears. $ T_{cl}$  is the mean temperature in the time interval  $\Delta t_{nucl}$  of the nucleating clusters
and $ n_{co.no.}$ is the mean coordination number of the particles at the centre of mass of the clusters and with mean potential
energy $u_{pot}$.
Data for the onset of nucleation are collected in Table 2. The data are the average values over the time interval
$\Delta t_{onset}=50$ (5000 time steps) at the beginning of the nucleation.
The  corresponding  data for the end of the nucleation time, $\Delta t_{end}=50$,  are also determined (Table 2, column 3).
$ $\\

$ $\\
$ $\\

{\bf Table 1}. Data for the cluster nucleation at $T=0.80$.
$ $\\
 \begin{tabbing} 
 \hspace{2.4cm}\= \hspace{2.9 cm}\=\hspace{2.4 cm}\=\hspace{2.4 cm}\=\hspace{2.5 cm}\=\hspace{2.5 cm}\kill
$\rho$\> $t_{nucl. start}$ \> $ \Delta t_{nucl}$  \> $ T_{cl}$ \> $ n_{co.no.}$  \> $u_{pot}$\\
 \end{tabbing}
 \begin{tabbing}
 \hspace{2.4 cm}\= \hspace{2.9 cm}\=\hspace{2.4 cm}\=\hspace{2.4 cm}\=\hspace{2 cm} \=\hspace{2 cm} \kill
0.027 \> 57000$_{\pm 44000}$ \> 343$_{\pm 276 }$ \> 0.782$_{\pm 9}$ \> 7.0$_{\pm 4}$ \>  -3.62$_{\pm 20}$ \\
0.0275 \> 46000$_{\pm 58000}$  \> 401$_{\pm 261}$  \> 0.782$_{\pm 11}$ \> 7.2$_{\pm 2}$ \>  -3.69$_{\pm 12}$ \\
0.028 \> 37000$_{\pm 26000}$  \> 390$_{\pm 225}$  \> 0.786$_{\pm 7}$ \> 7.2$_{\pm 4}$ \>  -3.69$_{\pm 21}$ \\
0.0285 \> 8800$_{\pm 8000}$  \> 359$_{\pm 203}$  \> 0.792$_{\pm 8}$ \> 6.9$_{\pm 4}$ \>  -3.59$_{\pm 20}$ \\
0.029 \> 8300$_{\pm 7800}$  \> 390$_{\pm 225}$  \> 0.795$_{\pm 9}$ \> 7.0$_{\pm 3}$ \>  -3.62$_{\pm 17}$ \\
 \>   \>   \>  \> $\rho \approx 0.60 $ \>
 \end{tabbing}
$ $\\
$ $\\

{\bf Table 2}. The mean temperature at the onset and at the end of nucleation.
$ $\\

 \begin{tabbing}
 \hspace{2.4 cm}\=\hspace{2.4 cm}\= \hspace{2.5 cm}\=\hspace{2.5 cm}\kill
$\rho$\> $S$\>  Onset: $ T_{cl} $ \>  End: $ T_{cl} $\\
 \end{tabbing}
 \begin{tabbing}
 \hspace{2.4 cm}\=\hspace{2.4 cm} \= \hspace{2.5 cm}\=\hspace{2.5 cm}\kill
0.027  \> 1.55 \>  0.765$_{\pm 18}$ \>0.789$_{\pm 19}$  \\
0.0275   \> 1.58 \>  0.771$_{\pm 20}$ \> 0.796$_{\pm 16}$ \\
0.028 \> 1.61 \> 0.776$_{\pm 17}$ \>0.797$_{\pm 16}$ \\
0.0285 \> 1.64   \> 0.783$_{\pm 17}$ \> 0.799$_{\pm 16}$   \\
0.029 \> 1.67  \> 0.788$_{\pm 16}$ \> 0.798$_{\pm 16}$  \\
 \end{tabbing}

$ $\\
$ $\\

The mean times, $t_{nucl. start}$,  for the start of nucleation (Table 1, Column 2) increase with decreasing
  QES densities and corresponding supersaturation $S$. The uncertainties of $t_{nucl. start}$, which are of the same order as $t_{nucl. start}$,
reflect the fact that homogeneous nucleating is a stochastic process. The mean nucleation times, $\Delta t_{nucl}$ (Column 3) are
$\approx 400$, corresponding to 40000 time steps. The temperature is obtained at each time step
during the nucleation.   The mean temperatures $T_{cl}$ (Column 4) are the mean of 25 independent nucleations
with standard deviations obtained from the 25 events, each with of the order 40000 instant
 values of the temperature during the nucleation. The mean temperatures
 are less than the mean
temperature  $T=0.80$ in the  systems, and there is  a systematic trend with colder nucleating clusters
for smaller  QES densities. The mean temperatures at the onset of nucleation (Table 2) confirm
this observation. The temperatures at the onset of nucleation are  colder than the mean temperature at nucleation, and colder with decreasing
QES densities. The mean temperatures at the end of nucleation  show that the
clusters now  are just colder than the supercooled gas, although not very much.\\
\begin{figure}[!]
\begin{center}
\includegraphics[width=7.5cm,angle=0]{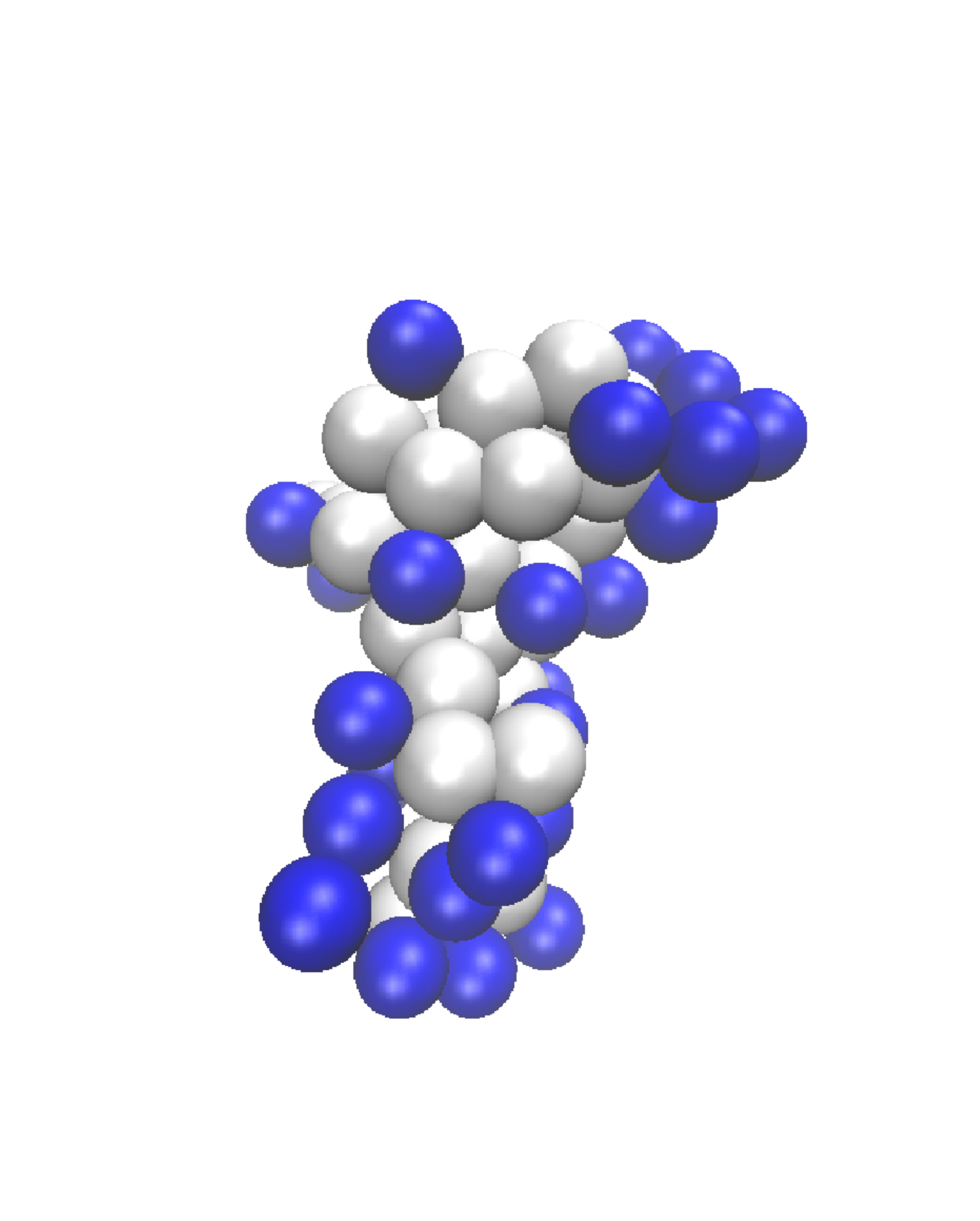}
\caption{ The  critical nucleus of $n_{cr}=70$ particles and with 33 liquid-like particles (grey). The instant temperature is
$T_{cl}=0.78$. The liquid-like core of the nucleus is not a compact spherical-like body.
}
\end{center}
\end{figure}

A question arises whether the low temperatures are caused by aggregations of cold particles and small cold clusters to the
surface of the growing nucleus or whether also the core of the clusters remain  colder than the mean temperature
in the QES.
 The core of the clusters are characterized by that the particles there have at least 
five  neighbours. The mean temperature in these cores
 for the 25 nucleations at $\rho=0.0275$ is $ T_{cl}=0.784_{\pm 8}$, and
in excellent agreement with the corresponding mean temperature $ T_{cl}=0.782_{\pm 11}$ for the clusters.
 So the cold  nucleating clusters are in internal
temperature equilibrium.   

 Another observation is that the mean lengths of the nucleation period, $\Delta t_{nucl}$ (Table1, Column 3) 
are almost the same for the
different state QES states despite of that the nucleation times varies with a factor of eight. 
The states with less QES densities overcome the transition barrier within the same time,
 but it is  only a  colder cluster which
  nucleates.

The mean coordination number, $n_{coor. no.}$, for the central particle in the clusters is smaller than the mean coordination
number for a particle $n_{coor. no.}(l)=9.02$ in the corresponding liquid. A mean coordination number in the interval 
$n_{coor. no} \in [6.9,7.2]$ (Table 1, Column 5) corresponds to a density $\rho \approx 0.60$. At the onset of nucleation
$n_{coor. no} \approx 6$, which corresponds to a center density $\rho \approx 0.53$ and at the end
of the nucleation period where $n_{coor. no} \approx 7.7$ the  density at the center of the critical nucleus is only
increased to  $\rho \approx 0.65$, which is  significant less than the liquid density  $\rho_l = 0.7367$ .
So the  nucleating cluster and the critical nucleus do not have  a  liquid-like center. This observation agrees with previous investigations \cite{Angelil}.
The mean potential energy per particle at the center of the nuclei is significant higher than the mean potential energy of corresponding bulk liquid
$u_{pot}(l)=-5.116$ due to the less dense core of the nucleating nuclei.

\begin{figure}[!]
\begin{center}
\includegraphics[width=8.5cm,angle=0]{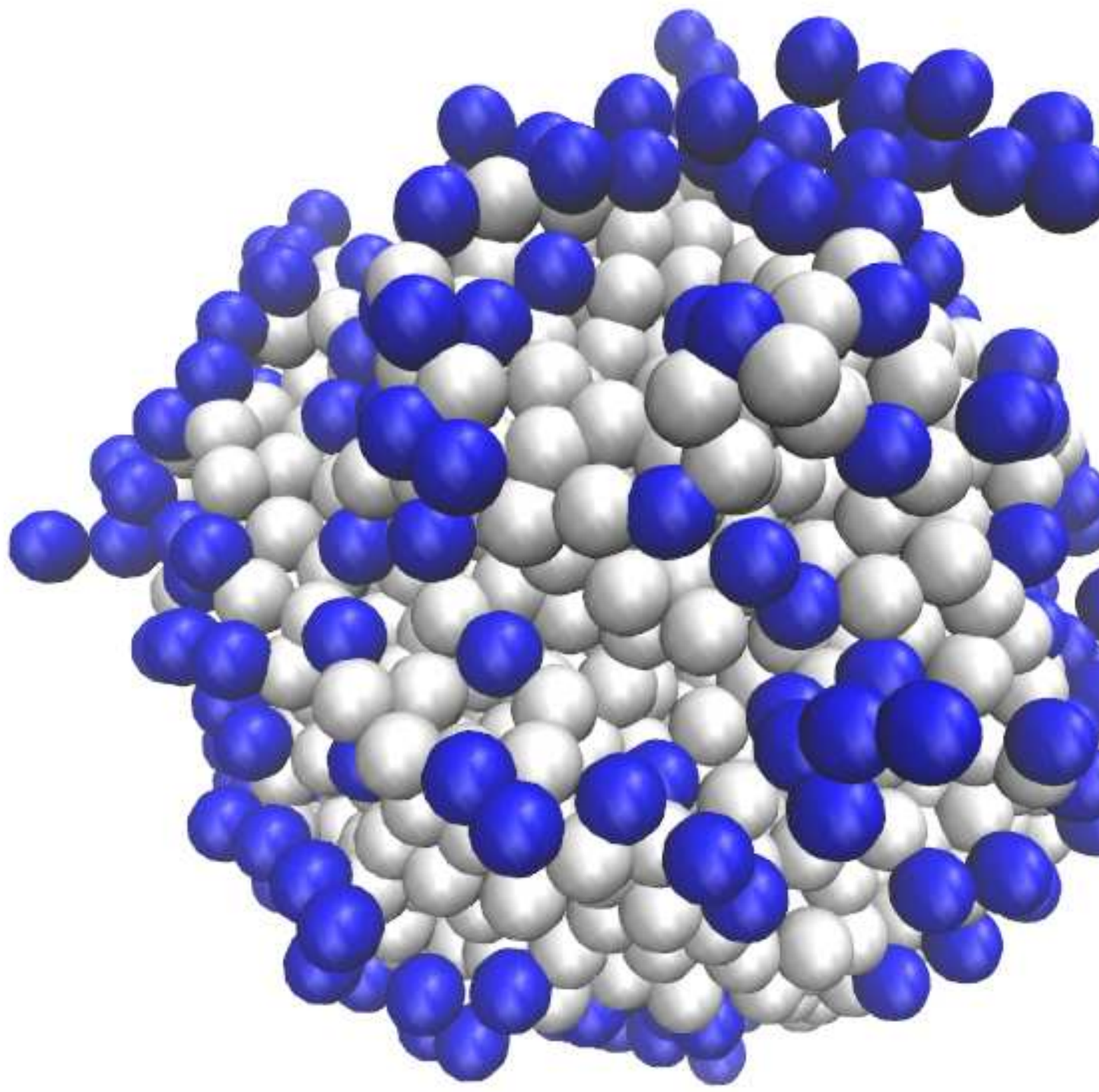}
\end{center}
\caption{ The nucleus after a time $t=1000$ with droplet growth. The nucleus now consists of 1665 liquid-like particles (grey) and  313
more weakly attached particles (blue).
}
\end{figure}

The critical nuclei  remain with a nucleus size of $n_{cl} \approx 70$ particles
 for a certain (transition) time in  many of the nucleations and for all five QES states (Figure 1).
This  "barrier tumbling" is because that
there is only a  small gain of free energy  by the growth
 at the extremum of the free energy. It is possible to locate a transition
zone in most of the nucleations.  The end   of the transition period  is
determined as  the time from where the size of the nucleus remain  above $n_{cr}=70$, and twenty four
 out of the twenty five nucleations for the QES with $\rho=0.0275$ 
 exhibit a transition zone, $\Delta t_{transition} > 10$.

 The behaviour in the transition zone  is
  investigated in details for the QES state       $\rho= 0.0275$. The mean temperature is now $T_{cl} =0.81_{\pm 1 }$ and equal to
the mean temperature in the system, but the mean coordination number $n_{co.no.}= 8.12_{\pm 1} \approx \rho= 0.68$ and potential
energy  $u_{pot} =-4.26_{\pm 9}$ of the central particle in the nuclei show that the critical nuclei in the transition zone
still do not have a liquid-like core. The liquid like core is however established rather short after the passage of the
critical nucleating barrier. Figure 5 shows  the nucleus after a time of $t=1000$ with droplet growth. The nucleus 
consist of 1665 liquid-like particles in a spherical droplet and with a bulk liquid core.

\section{Summary}
Homogeneous nucleation in a supercoled gas of Lennard-Jones particles is obtained by  MD simulations. 
The temperature  of the  nucleating cluster  is for the first time obtained every time step, and the simulations reveal that
it   is always a
  small cold cluster which initiates the nucleation. The temperature
in a nucleating cluster fluctuates, but the mean
temperature of the cluster remains below the temperature in the supersaturated gas until the critical nucleus is created. 
The temperature of the nucleating cluster changes in a systematic way with the degree of supersaturation.
 For less supersaturation  an even colder cluster starts the nucleation (Table 1,Table 2).
The  initial results are obtained for five moderate supersaturations at a  temperature $T=0.80$
 between the triple- and critical temperature. We have, however, tested that the dynamics are 
 valid  for  nucleation of liquid droplets at other temperatures ($T=0.70$ and 0.90), but further work needs to be carried out to
clarify e.g. whether it also is valid for more complex systems like nucleation of water droplets.

The critical nucleus has a temperature $T \approx 0.80$,
  equal to the temperature in the  gas   and with a  slow growth,
indicating that the critical nucleus is in an extremum   state of the free energy.
The critical nuclei do not have a compact  spherical shape, and the coordination number of the central
particles in the critical nuclei show that  the density at their centre
are not liquid-like, as also observed by \cite{Angelil}.
 A representative example is shown in Figure 4. The  nuclei take,
 however, quickly  a compact spherical-like form during the  succeeding droplet growth (Figure 5).
 
The kinetics of homogeneous nucleation, $\delta n_{cl}(t)$, is not only caused
by a grow or shrink by accretion or evaporation of
 monomers only, but by an exponentially declining
change in cluster size per time step.
 The kinetics at nucleation
 is  equal to the kinetics in the supercooled quasi equilibrium state, and given by the cluster distribution
in the supercooled gas (Figure 2), as expected for the kinetics  in an (quasi-) equilibrium state.

\section*{Acknowledgments}
The center for viscous liquid dynamics ‘Glass and Time’
is sponsored by the Danish National Research Foundation
(DNRF).


\begin{thebibliography}{99}
\bibitem{ToxMD} In MD the mass $m$ is usually included in the time unit.
 The unit length, energy and time used for LJ systems are, respectively,
 $\sigma$, $\epsilon$ and $\sigma\sqrt{m/\epsilon}$.
\end{thebibliography}
\end{document}